# Compact sub-Hz Linewidth Laser Enabled by Self Injection Lock To a Sub-mL FP Cavity

WEI LIANG*, YUNFENG LIU

*Suzhou Institute of Nano-Tech and Nano-Bionics (SINANO), Chinese Academy of Sciences, Suzhou 215123, China*
*\*Corresponding author: wliang2019@sinano.ac.cn*

**Narrow linewidth laser(NLL) of high frequency stability and small form factor is essential to enable applications in long range sensing, quantum information and atomic clocks. Various high performance NLL have been demonstrated by Pound-Drever Hall(PDH) lock or self injection lock(SIL) of a seed laser to a vaccum-stabilized FP cavity of ultrahigh quality factor(Q). However they are often complicated lab setups due to the sophisticated stabilizing system and locking electronics. Here we report a compact NLL of 68mL volume, realized by SIL of a diode laser to a miniature FP cavity of $7.7 \times 10^8$ Q and 0.5mL volume, bypassing table-size vaccum, thermal and vibration isolation. We characterized the NLL with a self-delayed heterodyne system, the Lorentzian linewidth reaches 60mHz, and the integrated linewidth is ~80Hz. The frequency noise performance exceeds that of commercial NLLs and the best reported hybrid-integrated NLL realized by SIL to high Q on-chip ring resonators. Our work marks a major step toward a field-deployable NLL of superior performance utilizing ultra-high Q FP cavity.**

Ultra Narrow linewidth Lasers(NLL) of high frequency stability and low size, weight and power(SWaP) have found wide applications in long range sensing, coherent communication, quantum information, high-resolution spectroscopy and atomic clocks[1-8]. In the past, fiber distributed feedback lasers and non-planar ring oscillator (NPRO) solid state lasers have met the market need, however they have relative large size, high power consumption and are limited to few wavelength selections. To meet the needs of applications requiring low SWaP, a number of external cavity diode laser (ECDL) products have been developed and welcomed by the market in the past decade. These ECDLs are realized by coupling of gainchip to narrow band on-chip reflective Bragg grating[9], or self injection lock(SIL) of DFB to micro-resonators of high Q factor[10-11]. In recent years, hybrid integrated narrow linewidth lasers(HI-NLL) realized by coupling diode laser chip to high Q on-chip ring resonator have attracted intensive studies and made rapid progress [12-15]. E.g. in a recent report the noise of HI-NLL already beats commercially available fiber laser[16]. Meanwhile, it has also been pointed out that the low offset frequency noise of the HI-NLL are mainly dominated by the thermal refractive noise of the waveguide based resonator due to the tight confinement of the light and the small modal volume, and hence increasing the modal volume by e.g. increasing the cavity length to 1.41 meter have been resorted to further reduce the low-offset frequency noise[16]. Nevertheless, a on-chip resonator of such large size makes it difficult to SIL to a mode of specific wavelength due to the very small FSR(~135MHz) and the wide pulling range of SIL(typically a few GHz).

By contrast, thanks to the very low absorption and thermal refractive noise, and the use of low-thermal expansion material, hollow FP cavity has remained to be the quietest optic cavity of all kinds. A carefully fabricated FP cavity offers finesse and short-term stability that are still orders of magnitude ahead of the best on-chip ring resonators. In the past, carefully stabilized FP cavities with volumes of a few Litters, have been used to demonstrate ~mHz linewidth lasers with Pound-Driver-Hall(PDH) lock technique[17], and 10Hz linewidth laser with SIL technique[18]. However these ultra-stable FP cavities require vaccum chamber and sophisticated thermal and vibration shield, which is difficult to deploy in the field. In a recent report of HI-NLL, the authors have also resorted to a µ-FP cavity of ~8mL volume, and achieved remarkably low frequency noise [19]. The system starts with a seed NLL composed of a gain chip coupled to a cm long surface grating, which is SILed to a on-chip high Q spiral resonator of 1.41 meter length. Further the seed NLL is PDH locked to the vacuum stabilized µ-FP cavity using acousto-optic modulator(AOM) and electric-optic modulator(EOM). Despite the excellent performance, the cavity stabilizing system and the PDH lock electronics make it still far from being a portable and affordable NLL.

In terms of locking a chip-scale diode laser to a microcavity, SIL is a lot simpler and more economic to apply compared with PDH lock. With the goal of providing industry a NLL of high performance and low SWaP, we marry the the simplicity of SIL with the low noise feature of a miniature FP cavity, and report a compact NLL enabled solely by SIL of a DFB laser to a high Q FP cavity of sub-mL size. The Lorentzian linewidth of the laser reaches 60mHz, and the integrated linewidth is approximately 80Hz. By beating two similar lasers we also measured the frequency drift at the level of $10^{-12}$ from 0.1ms to 100ms, and it stays below $10^{-11}$@1s if the linear drift is removed. The frequency noise and stability beat commercially available ECDL NLLs and is way ahead of state-of-art HI-NLL using solely SIL to on-chip ring cavity[16,19]. Of note, compared with the HI-NLL PDH locked to a stabilized FP cavity[19], our laser's frequency noise at off-set frequency higher than 100kHz is 10dB lower, thanks to the much higher noise reduction bandwidth offered by SIL technique. This demonstration marks a significant step toward a field-deployable NLL utilizing hollow FP cavity of ultra-high Q value.

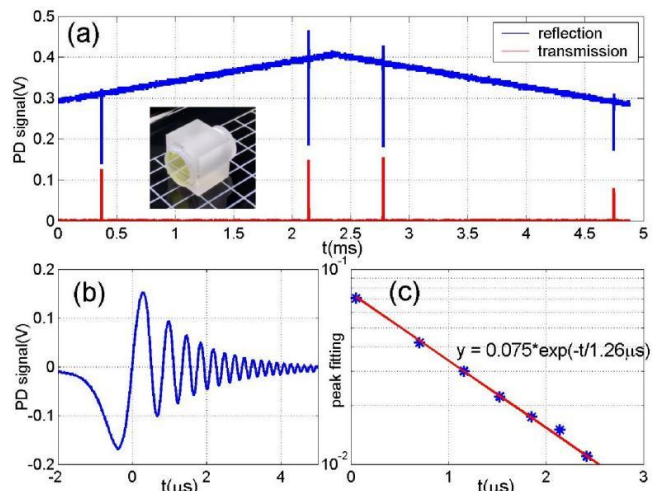

Fig.1. Characterization of the miniature FP cavity. (a) Measured transmission(red curve) and reflection(blue curve) spectrum of the FP cavity using a DFB laser. The DFB laser's frequency is swept more than one FSR(~21GHz) of the cavity with a period of 5ms and the signals are captured by fast PDs on an oscilloscope. The inset is a picture of the actual resonator sitting on a Gel-Pak box. The size of the cavity spacer is 8x8x7mm. (b). Ring down trace extracted from the reflection signal. (c). Fitting of the amplitude of the oscillating ring down signal. The time constant is 1.26μs and the induced Q value is ~770 million.

Our sub-mL high Q FP cavity is formed by attaching a planar mirror(diameter of 6.35mm, thickness of 1mm) and a concave mirror (diameter of 6.35mm, thickness of 2.3mm, radius of curvature of 250mm) to a 7mm long hollow fused silica spacer. The inset of Fig. 1a shows an actual resonator on a Gel-Pak box. The volume of the cavity is slightly less than 0.5mL. Both mirrors are made of fused silica and have nominal reflectivity of 99.996%. The FSR is ~21GHz, and the theoretical linewidth is 240kHz ignoring the absorption and scattering loss of the mirrors. Although a FP cavity has many transverse modes, by carefully designing the input beam to match that of the fundamental mode, one can achieve a good side mode suppression ratio(SMSR). In our case we used a DFB source and a fiber collimator delivering a Gaussian beam of 350μm diameter to characterize the cavity. Fig. 1a shows the reflection and transmission spectrum of the cavity by sweeping the current of the DFB laser, and high-order transverse mode is almost invisible on the oscilloscope indicating an excellent mode match to the fundamental mode. Fig. 1b shows the ring-down trace extracted from the reflection signal, the fitted decaying time constant is 1.26μs and the bandwidth of the cavity is around 252kHz [20], which is very close to the theoretical value. In addition we don't see obvious asymmetric mode shape typically observed in whispering gallery mode(WGM) and on-chip ring resonators caused by thermal non-linearity[21,22] even with pump power of tens of mW, which is an indication of the low absorption of the mirrors.

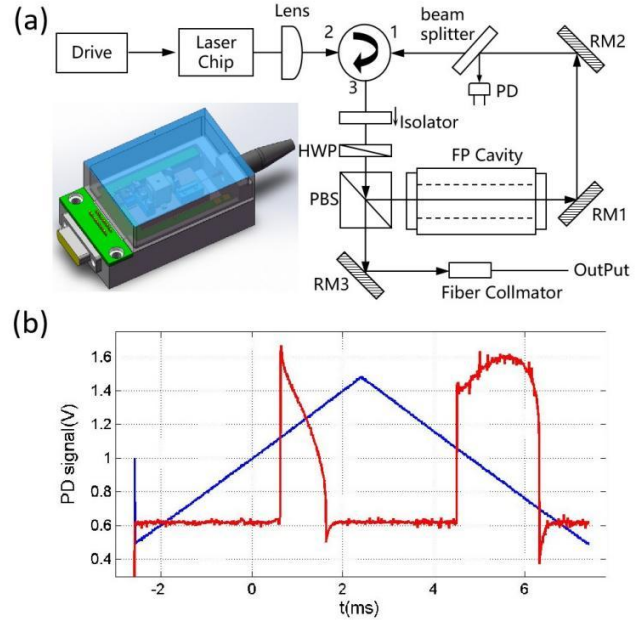

Fig. 2a. Schematic diagram of the NLL module. HWP: half wave plate. PBS: polarization beam splitter. RM: reflection mirror. The inset is an illustration of the NLL module. Fig 2b. Signal of PD monitoring the transmission of the cavity and the injection locking status of the DFB. The laser's driving current is swept by 35mA with a triangle wave corresponding to ~12GHz frequency ramp of the free running DFB. The PD signal of the right part of the triangle sweep indicates that the locking range is more than 4GHz with the appropriate feedback phase.

Fig. 2a shows the schematic diagram of the NLL laser. Emission of a 80mW DFB is collimated and passes through a freespace circulator, the reflected light is split by a cube beam splitter, and roughly 30% light is reflected and coupled toward the FP cavity, the other 70% light goes through an isolator and couples to a polarization maintaining(PM) fiber collimator with maximum output power >40mW. With up to 20mW power pumped into the cavity, we don't observe exotic nonlinear effects. The transmission of the FP cavity is routed by two mirrors and coupled back to the DFB through the circulator to complete the SIL. A small amount of light is split and routed toward a monitor PD to detect the cavity's transmission power and to monitor the SIL status. The inset is an illustration of the NLL laser module design. The total volume is 68mL and the optic bench containing all the optic elements has a volume of only 8 mL. We use a home-developed driver to provide low noise current to the DFB, and to stabilize the temperatures of the DFB and the optic bench respectively. The driving current is up to 300mA and the temperatures stability is better than 10 mK. Fig. 2b shows the typical PD signal as the frequency of the DFB is swept through the fundamental mode of the cavity. The triangle current sweep range is 35mA corresponding to ramp frequency range of ~12GHz of the free running DFB. Inside the 12GHz frequency sweep range we only observe SIL to the fundamental mode, indicating a good mode match between the collimated beam and the cavity fundamental mode is achieved. From the width of the locked PD signal of the right part of the triangle sweep, we can estimate that the locking range is more than 4GHz when the feedback phase is appropriate.

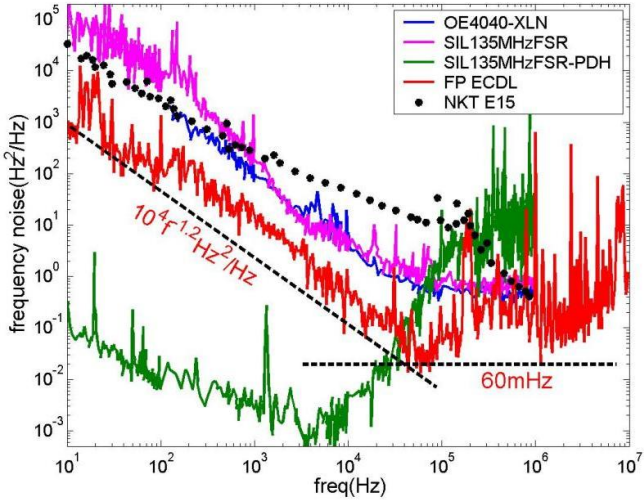

Fig. 3 Characterization of the frequency noise. The frequency noise of our narrow linewidth laser(red line) is measured with an AOM and a fiber interferometer of 1km differential delay according to [23]. The black dot curve is a typical frequency noise curve of NKT E15 fiber laser. The blue curve is the frequency noise of OE4040-XLN ECDL product. The purple and green lines are the frequency noise curves of the free running and PDH-locked HI-NLL, respectively[19].

To characterize the performance of the SIL laser, we select the DFB's current so that the PD signal stays close to the top of the red curve in Fig. 2b, which is close to the center of the cavity transmission. We performed the SSB frequency noise measurement using an AOM and a fiber interferometer of 1km differential delay according to [23]. In Fig. 3 we plot the frequency noise of our NLL(red line). The lorentzian linewidth given by the white frequency noise drops from ~100kHz(typical value of free running DFB) to ~60mHz with reduction factor ~$10^6$. If we draw a dashed line trending the floor of the measured frequency noise curve, it's described by $10^4 f^{-1.2}$ Hz$^2$/Hz, and the integrated linewidth defined according to [24] is approximately ~ 80Hz. In Fig 3 we also include the frequency noise curves of well-known commercial products e.g. NKT E15(black dots), and the recently reported HI-NLL(purple line) for the purpose of comparison. As one can see that the frequency noise of our NLL is lower at all off-set frequencies. It's also obvious the frequency noise of our laser is 30~40dB worse than the HI-NLL PDH locked to a stabilized μ-FP cavity at offset frequency <10kHz. This is understandable as the μ-FP cavity used in the work [19] has a linewidth ~10kHz which is ~15 times better than the μ-FP cavity used in our laser. In addition PDH lock always locks to the center of the mode and reduces the jitter caused by current and temperature fluctuation, hence offering tighter locking than SIL within the feedback bandwidth. Nevertheless at offset frequency >40kHz, our NLL has better frequency noise, thanks to the higher noise suppression bandwidth of SIL. This is particularly important for quantum sensors with observation frequencies expanding to higher frequency [25,26]. The spur and hump at frequency 10~100Hz are due to the residual vibration noise of the measurement fiber interferometer and power line noise of the laser driver.

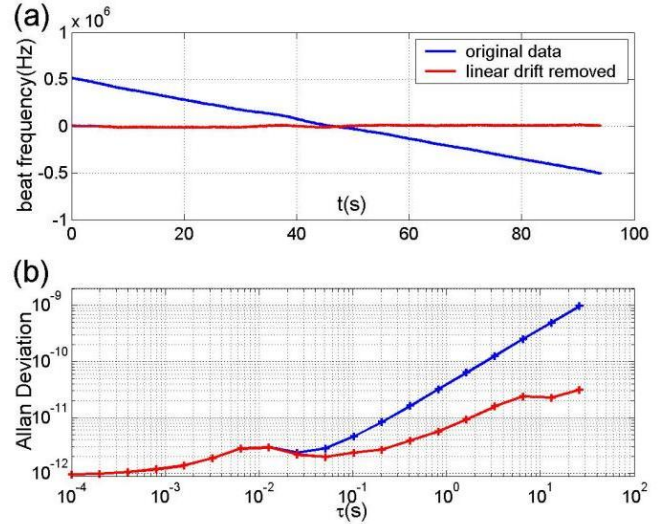

Fig. 4 Measurement of the laser's frequency stability. (a) The beat-note frequency of two similar narrow linewidth lasers is down converted and recorded by a frequency counter with gate time of 0.1ms. (b) calculated Allan Deviation. For both (a) and (b), the blue curve represents the original data and the red curve represents the results with the linear drift removed.

We further characterized the frequency drift of the laser by beating two similar lasers. The heterodyne beat signal is down converted to 170MHz from ~18GHz using a tunable signal source locked to a temperature Compensated crystal Oscillator(TCXO), and the down converted signal is counted by a Agilent frequency counter 53230A. Fig. 4(b) shows a typical measured frequency drift curve and the calculated allan deviation(AD), the AD of a single laser is around $10^{-12}$ from 1ms to 100ms, and then picks up at slope of which is dominated by the linear drift. It follows the trend of random walk and is still below $10^{-11}$ @1s if the linear drift is removed. In addition the hump at 100ms can be traced to the power line noise(50Hz) affecting the laser driver.

In summary, we have fabricated a μ-FP cavity with quality factor of 770 million and volume of slightly less than 0.5mL. We also packaged compact NLLs realized by SIL of a 80mW DFB laser to the μ-FP cavity. The FP cavity and the laser are temperature controlled with TECs and home-made driver, with neither vacuum vessel nor vibration isolation needed. The small volume of the FP cavity and the simple locking scheme allows us to package the system with a total volume of 68mL. In fact the volume of the optic bench is merely 8 mL and it's possible to further shrink the module to fit into a butterfly package. This represents a substantial improvement in terms of form factor and portability compared with NLLs previously demonstrated with high Q hollow FP cavity. Furthermore, the large modal volume combined with the very low nonlinearity of the μ-FP cavity allows us to use high power pump laser with no need to reduce the Q factor of the cavity and sacrifice the resulted noise performance. Looking into future, we expect to further reduce the lasers' frequency noise by orders of magnitude, if a μ-FP cavity with Q factor of more than 10 billion and ULE spacer is adopted. Finally we want to stress our NLL design can be mass-produced with modern telecom optic module assembly line, and can also be expanded to other wavelengths for various demanding clock and quantum applications outside the laboratory.

**Funding.** This work was supported by the National Natural Science Foundation of China under Grant 011010201.

**Acknowledgments.** The authors also thanks the Chinese Academy of Science, Jiangsu Province, SuZhou and Suzhou Industrial Park for the support of the work.

**Author contribution.** Wei Liang conceived the idea and the optical layout, performed the measurements, conducted the data analysis and wrote the manuscript. Yunfeng Liu performed the mechanical design and assembled the narrow linewidth lasers.

**Disclosures**. The authors declare no conflicts of interest.

**Data availability.** Data underlying the results presented in this paper are not publicly available at this time but may be obtained from the authors upon reasonable request.

## References

1. J.C. Juarez, E.W. Maier, Kyoo Nam Choi, H.F. Taylor, J. Lightwave Technol. 23, 2081-2087, 2005.
2. Y. Koshikiya, X. Fan, and F. Ito, J. Lightwave Technol. 28, 3323-3328, 2010.
3. G. M. Brodnik, M. W. Harrington, J. H. Dallyn, D. Bose, W. Zhang, L. Stern, P. A. Morton, R. O. Behunin, S. B. Papp, and D. J. Blumenthal, Nat. Photonics 15(8), 588-593 (2021).
4. M. Lucamarini, Z. L. Yuan, J. F. Dynes, and A. J. Shields, Nature 557(7705), 400-403 (2018).
5. M. Jing, Y. Hu, J. Ma, H. Zhang, L. Zhang, L. Xiao, and S. Jia, Nat. Phys. 16(9), 911-915 (2020).
6. A. Chopinaud and J. D. Pritchard, Phys. Rev. Appl. 16(2), 024008 (2021).
7. R. J. Niffenegger, J. Stuart, C. Sorace-Agaskar, D. Kharas, S. Bramhavar, C. D. Bruzewicz, W. Loh, R. T. Maxson, R. McConnell, D. Reens, G. N. West, J. M. Sage, and J. Chiaverini, Nature 586(7830), 538-542 (2020).
8. Y. Jiang, A. Ludlow, N. Lemke, R. Fox, J. Sherman, L.-S. Ma, and C. Oates, Nat. Photonics 5(3), 158-161 (2011).
9. K. Numata, J. Camp, M. A. Krainak, and L. Stolpner, Opt. Express 18(22), 22781–22788 (2010)
10. W. Liang, V. S. Ilchenko, A. A. Savchenkov, A. B. Matsko, D. Seidel, and L. Maleki, Opt. Lett. 35(16), 2822-2824 (2010).
11. W. Liang, V. Ilchenko, D. Eliyahu, A. Savchenkov, A. Matsko, D. Seidel, and L. Maleki, Nat. Commun. 6(1), 7371 (2015).
12. B. Stern, X. Ji, A. Dutt, and M. Lipson, Opt. Lett. 42(21), 4541-4544 (2017).
13. M. A. Tran, D. Huang, and J. E. Bowers, APL Photonics 4(11), 111101 (2019).
14. W. Jin, Q. F. Yang, L. Chang, B. Q. Shen, H. M. Wang, M. A. Leal, L. Wu, M. D. Gao, A. Feshali, M. Paniccia, K. J. Vahala, and J. E. Bowers, Nat. Photonics 15(5), 346-353 (2021).
15. C. Xiang, J. Guo, W. Jin, L. Wu, J. Peters, W. Xie, L. Chang, B. Shen, H. Wang, Q.-F. Yang, D. Kinghorn, M. Paniccia, K. J. Vahala, P. A. Morton, and J. E. Bowers, Nat. Commun. 12(1), 6650 (2021).
16. B. Li, W. Jin, L. Wu, L. Chang, H. Wang, B. Shen, Z. Yuan, A. Feshali, M. Paniccia, K. J. Vahala, and J. E. Bowers, Opt. Lett. 46(20), 5201–5204 (2021)
17. A. D. Ludlow, X. Huang, M. Notcutt, T. Zanon-Willette, S. M. Foreman, M. M. Boyd, S. Blatt, and J. Ye, Opt. Lett., 32(6), 641-643, 2007.
18. Y. Zhao, Y. Li, Q. Wang, F. Meng, Y. G. Lin, S. K. Wang, B. K. Lin, S. Y. Cao, J. P. Cao, Z. J. Fang, T. C. Li, and E. J. Zang, IEEE Photon. Technol. Lett. 24(20), 1795-1798 (2012).
19. J. Guo, C. A. Mclemore, C. Xiang, D. Lee, L. Wu, W. Jin, M. Kelleher, N. Jin, D. Mason, L. Chang, A. Feshali, M. Paniccia, P. T. Rakich, K. J. Vahala, S. A. Diddams, F. Quinlan, J. E. Bowers, arXiv:2203.16739 [physics.optics]
20. G. Rempe, R. J. Thompson, H. J. Kimble, and R. Lalezari, Opt. Lett. 17(5), 363-365 (1992).
21. V. S. Il'chenko and M. L. Gorodetskii, Laser Phys. 2, 1004-1009 (1992).
22. X. Jiang and L. Yang, Light: Sci. Appl. 9(1), 2047-7538 (2020).
23. S. Camatel and V. Ferrero. J. Lightwave Technol. 26(17), 3048 (2008).
24. D. R. Hjelme, A. R. Mickelson, R. G. Beausoleil, and J. A. McGarvey, IEEE J. Quantum Electron 27, 352-372 (1991)
25. R. Li, F. N. Baynes, A. N. Luiten, and C. Perrella, Phys. Rev. Appl. 14(6), 064067 (2020)
26. M. Jing, Y. Hu, J. Ma, H. Zhang, L. Zhang, L. Xiao, and S. Jia, Nat. Phys. 16(9), 911-915 (2020).